\documentclass[twocolumn,showpacs,preprintnumbers,amsmath,amssymb]{revtex4}
\usepackage{graphicx}
\usepackage{amsmath,amsfonts,amssymb}  
\usepackage{dcolumn}
\usepackage{bm}

\begin{document}

\title{On Deviations from Gaussian Statistics for Surface Gravity Waves}
\author{M. Onorato, A. R. Osborne, M. Serio}
\affiliation{Dip. di Fisica Generale, Universit\`{a} di Torino, via
Pietro Giuria 1, 10125 Torino, Italy}
\date{\today}

\begin{abstract} 
Here we discuss some issues concerning 
the statistical properties of ocean surface waves.
We show that, using the approach 
of weak turbulence theory, deviations from Gaussian statistics
can be naturally included. In particular
we  discuss the role of 
bound  and free modes for the determination 
of the statistical properties of the surface elevation. 
General formulas for skewness and kurtosis 
as a function of the spectral wave action density
are here derived.
\end{abstract}

\maketitle
\section{Introduction}

Recently it has been observed 
experimentally that large amplitude waves 
can appear on the surface of the ocean. There are
different mechanisms that can lead to the formation 
of such events. For example, linear theory can 
be considered as a first candidate for explaining 
these waves: it can happen
that for some fortuitous occasion, two or more 
waves of different lengths are in phase, leading to 
the well known constructive interference mechanism 
 (linear superposition of Fourier modes). 
When the surface elevation is characterized 
by a large number of linear
waves with random phases, it is  
possible to estimate the 
probability of measuring an extreme
event (say a wave of height larger than 2 times
the significant wave height). This theoretical
task has been accomplished about 50 years ago 
\cite{LH52} 
and the main 
result is that, if the surface elevation 
is Gaussian and the process is narrow-banded, 
then wave heights and wave crests are distributed according
to the Rayleigh distribution (corrections due to finite 
spectral band-width have also been obtained). 

First substantial corrections to this distribution can 
be obtained if waves are considered weakly nonlinear, or 
more precisely, if for each wave component (free modes),
 its bound contributions are included. For the 
narrow-banded case this is nothing but 
describing the surface elevation as a Stokes 
expansion.
The more general description of the surface elevation,
valid for any spectral band-width, was given in a seminal 
paper by Longuet-Higgins \cite{LH63}. He was 
able to derive the contributions to 
the surface elvation from bound waves 
 up to second order in wave steepness. 
The numerical implementation of the formulas reported in 
the paper by Longuet-Higgins corresponds to what today 
is called a ``second order theory'' 
(see \cite{FOR00}); 
note that the theory includes only bound modes and not free modes. 
The presence of those Stokes-like contributions gives the
 waves the well known property of being positively skewed.
Using the second order theory, \cite{TAY80}
  was able to include this
contribution in the  distribution function of the 
wave crests. More in particular, for unidirectional
narrow-banded waves in infinite water depth, with the 
hypothesis that free waves are described by 
a Gaussian statistics,  he derived a formula
for the distribution of wave crests (now known as the Tayfun 
distribution) which 
does bring substantial corrections to the 
Rayleigh distribution, especially if the wave steepness 
is large. It should be here
stressed that the Tayfun second order theory predicts a Rayleigh 
distribution for wave heights 
(this is because second order
contributions cancel out for  wave heights).
The Tayfun distribution has been recently 
compared successfully with interesting numerical experiments 
of envelope equations
(see  \cite{SO05}).
Paradoxically, the distribution describes very well the
data characterized by large directional spreading but underestimates
wave crests in the long-crested case for which 
the distribution has been derived.

Only in the last few years it was realized 
 that not only bound modes can generate deviations
from a Gaussian statistics but also the dynamics
of free waves should be considered in the determination of
the statistical properties of the surface elevation.
More in particular, it was shown in 
\cite{ONO01} and 
\cite{JAN03} that the nonlinear interactions
of free modes can substantially alter the statistical 
properties of the surface elevation. Note that in this
case the statistics of free mode
 (without the contribution from bound modes) can be non Gaussian.
It was also found that the nonlinear interactions
responsible for such a deviation from 
gaussian statistics are  associated
with the modulational instability mechanism (also known as the 
Benjami-Feir instability) which can be thought as
a quasi-resonant 4 wave interaction. Those concepts
are the bases of the theory developed in \cite{JAN03},
where a kinetic equation that includes quasi-resonant interactions
is derived. More than that, in the same paper, an 
equation that relates the kurtosis
(fourth moment of the surface elevation) to the spectral
wave action density is obtained. The theory includes only
the contribution to the kurtosis from free modes.

The aim of the present paper is to describe a single theory,
based on the Hamiltonian formulation of surface gravity 
waves, that can take into account both 
the contribution of bound 
and free waves. Before entering in the discussion 
we will motivate the importance of including the 
free wave dynamics in the theory by
showing some experimental data
recorded during the  Marintek experiments
\cite{ONO04}. Here we anticipate that 
the experimental data suggest that
in the particular condition of long crested waves and 
large Benjamin-Feir Index, 
the second order theory is inadequate to describe the 
distribution of wave crests. 
In the last part of the paper we will sketch the derivation 
of the formulas for the kurtosis and skewness which 
include both the contribution from bound and free modes.

\section{Data from Marintek compared with the Second Order Theory}
Here we will consider some experimental data recorded 
at Marintek in one of the largest wave
tank in the world. 
 The length of the tank is $270$ $m$
and its width is $10.5$ $m$.
The conditions
at the wave maker were provided by a Jonswap spectrum
with random phases. One accepts this fact and then 
lets the phases (and amplitudes) evolve according to the 
nonlinear dynamics. Different runs were performed (see
\cite{ONO04}). 
Here we will consider just the probability 
density function of wave crests for the run characterized by a strong nonlinearity (steepness 
calculated as $\epsilon=H_s k_p/2$ was about 0.15, 
with $H_s$ the significant
wave height and $k_p$ the wave-number of the peak of the spectrum, computed 
using the linear dispersion relation from the peak frequency).
Experimental data are compared with numerical data from 
a standard second order theory (the coupling coefficient
has been taken from  \cite{FOR00}
 and with the Tayfun  distribution. 
 
\begin{figure}
\centerline{\includegraphics[width=0.5\textwidth]{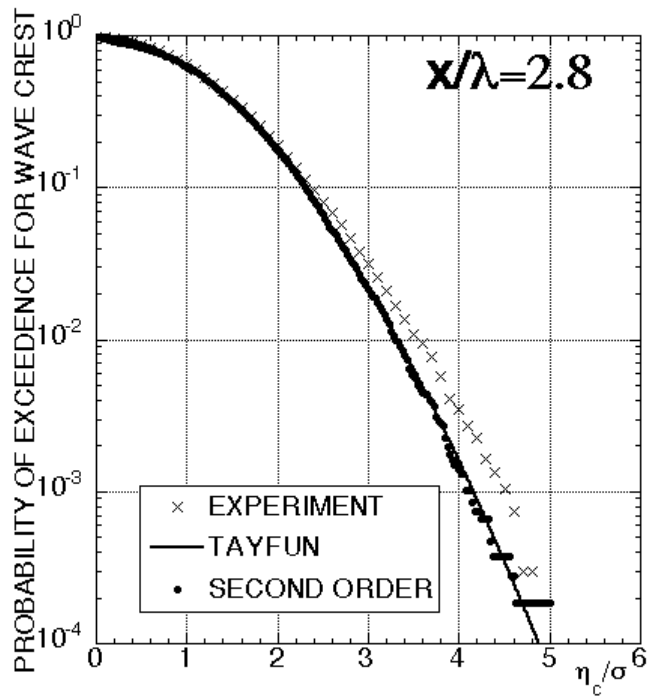}}
\caption{Wave crest distribution: experimental 
  data, measured at 2.8 wavelengths from the wave maker, 
  are compared with data from Tayfun and 
  second order theory.}
  \label{fig:figura1}
\end{figure}

In figure ~\ref{fig:figura1} we show a comparison between 
experimental the wave crest distribution 
recorded at the first probe, a few wave lengths from
the wave maker, and the second order theory.
First of all, it should be mentioned that for the present conditions
the Tayfun theory  
is in good agreement with second order theory; the experimental 
data are not so far from the mentioned
distributions. 
\begin{figure}
\centerline{\includegraphics[width=0.5\textwidth]{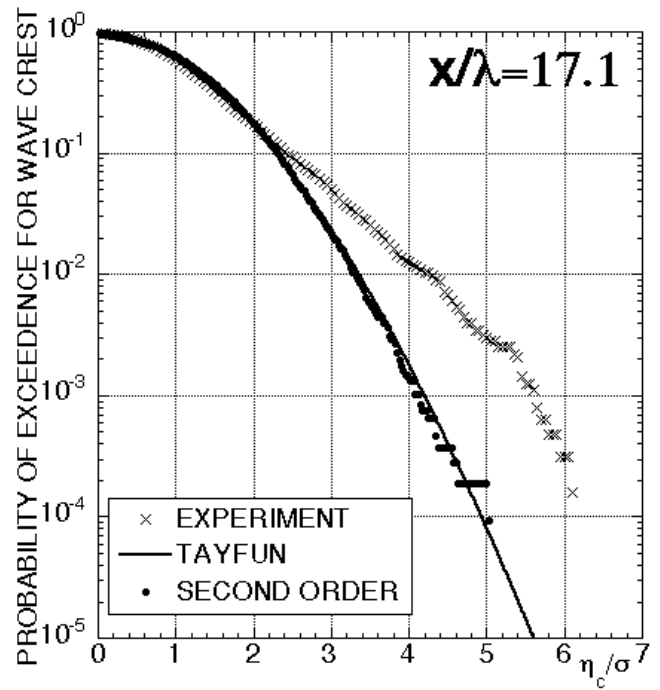}}
\caption{Wave crest distribution: experimental 
  data, recorded at 17.1 wavelengths from the wave maker,
   are compared with data from Tayfun and 
  second order theory.}
  \label{fig:figura2}
\end{figure}
This is not a surprise because data have 
been built as a linear superposition of Fourier modes and 
the generation of bound modes 
is practically instantaneous with respect to 
the Benjamin-Feir space scales. The situation is 
different after 
the waves have evolved along the tank 
according to their nonlinear dynamics. As can be seen in 
figure ~\ref{fig:figura2}, the Tayfun distribution 
(and the second order theory) completely
underestimate  the experimental data. 
This result is quite significant 
because it clearly tells us that there are some statistically stationary 
conditions in which the second order theory is completely inadequate
to describe the probability density function of real water waves.

Apparently there is no reason for the second order theory 
to fail to reproduce
the experimental data; data (steepness and spectral 
shape considered) do not violate any of the assumption for deriving the 
second order theory. 
In the second order theory 
the surface elevation is considered 
 as a linear superposition of non interacting free 
modes to which  bound modes have been added.
The main problem in the approach is that, 
as it is well known from Hasselmann-Zakharov theory
(\cite{HAS62},  \cite{ZAK67}), 
 free waves also interact.
In the next section we will consider a general theory which 
considers both the contribution 
from bound modes and free modes.

\section{Skewness and Kurtosis: Contribution from Bound and Free Waves}
The goal of this section is to derive formulas for 
the skewness and the kurtosis as a function of 
the spectral  wave action density.
The starting point for the derivation of the theory is
the Hamiltonian description of surface gravity waves
(see \cite{ZAK68} and \cite{ZAK99}  ).
The theory described below is general and 
can be applied to any weakly nonlinear dispersive system.
 Here we will start with the following
 Hamiltonian, truncated to four-wave interactions:
\begin{equation} \label{HAMIL}
\begin{split}
&H=\int\omega_0 a_0 a_0^* dk_0 + \\
&+ \int U^{(1)}_{0,1,2} (a_0 a_1^* a_2^* +
a_0^* a_1 a_2) \delta_{0-1-2} dk_{012}+  \\
& + \frac{1} {3} \int U^{(2)}_{0,1,2} 
(a_0^* a_1^* a_2^* + a_0 a_1 a_2) \delta_{0+1+2} dk_{012} +\\
& + \frac{1} {2} \int V^{(2)}_{0,1,2,3} 
a_0^* a_1^* a_2 a_3 \delta_{0+1-2-3}dk_{0123} + ...
\end{split}
\end{equation}
The notation is taken from \cite{KRA94}
(for example, $a_i=a({\bf k}_i)$, 
$U^{(1)}_{0,1,2}=
U^{(1)}({\bf k}_0,{\bf k}_1,{\bf k}_2)$). 
The coupling coefficients
and their properties are reported in the just mentioned paper 
(the reader unfamiliar with the present Hamiltonian 
description of surface gravity waves who is willing to understand 
details of what follows is strongly advised to read the paper
\cite{KRA94}). Note that we have for brevity omitted 
three more integrals which contain non 
resonant four-wave interactions.
Up to our desired accuracy, those terms do not contribute
to the skewness and to the kurtosis,
therefore can be safely neglected.
The surface elevation $\eta({\bf k})$ and the velocity 
 potential $\psi({\bf k})$
are related to the wave action variable $a({\bf k})$ 
in the following  way (note that the dependence on 
time has been omitted for brevity):
\begin{equation} \label{SURF}
\eta({\bf k})=\sqrt {\frac {k} {2\omega({\bf k})}} 
(a({\bf k})+a^*(-{\bf k}))
\end{equation}
\begin{equation} \label{POT}
\psi({\bf k})=-i \sqrt {\frac {\omega({\bf k})} {2k}} 
(a({\bf k})-a^*(-{\bf k})).
\end{equation}
It is well known that three waves resonant interactions
are forbidden for surface gravity waves, therefore it is always possible to introduce 
 new canonical variables $b$ and $b^*$ for which the 
Hamiltonian does not show explicitly those interactions.
The standard way of doing this consists in considering the canonical 
transformation, expressed as the following infinite series: 
\begin{equation}\label{CAN}
\begin{split}
 a_0=b_0 + & \int A^{(1)}_{0,1,2}  b_1 b_2 \delta_{0-1-2} dk_{12}+ \\
&  +\int A^{(2)}_{0,1,2} b_1^* b_2 \delta_{0+1-2} dk_{12}+\\
& +\int A^{(3}_{0,1,2} b_1^* b_2^* \delta_{0+1+2} dk_{12}+...
\end{split}
\end{equation}
The canonical transformation allows one 
to separate directly the bound modes from 
the free modes: modes $b(\bf k)$ are free modes which have their
own dynamics and the surface elevation (which 
contains free and bound modes) 
can be recovered directly by using equations
 (\ref{SURF}) and (\ref{CAN}).
Indeed, it is easy to show that in
the narrow-banded approximation equations (\ref{SURF}) and
 (\ref{CAN}) 
 reduce to the Stokes expansion, i.e. a carrier
wave plus its bound harmonics.

Before entering into the details, here we mention 
that we will make large use of the so called
{\it quasi-gaussian approximation}, i.e. the goal is to 
express the higher order correlators
as a function of the second order correlator, 
$<b_1 b_2^*>=N_1\delta_{1-2}$, with $N=N({\bf k})$
the wave action density spectrum 
($<...>$ indicate ensemble averages). In this approximation
the fourth and six order correlators for the free
waves can be writen as follows (see for example 
\cite{JAN03})
\begin{equation} \label{CORRE4}
<b_1 b_2 b_3^* b_4^*>=N_1 N_2 (\delta_{1-3} \delta_{2-4}+
\delta_{1-4} \delta_{2-3})+D_{1,2,3,4}
\end{equation}
\begin{equation} \label{CORRE6}
\begin{split}
<b_1 b_2 b_3 b_3^* b_4^* b_5^*>& =N_1 N_2 N3\\
& (\delta_{1-4} \delta_{2-5} \delta_{3-6}+
\delta_{1-4} \delta_{2-6} \delta_{3-5}+\\
&+ \delta_{1-5} \delta_{2-4} \delta_{3-6}+
\delta_{1-5} \delta_{2-6} \delta_{3-4} +\\
&+\delta_{1-6} \delta_{2-4} \delta_{3-5}+
\delta_{1-6} \delta_{2-5} \delta_{3-4})
\end{split}
\end{equation}
$D_{1,2,3,4}$ is the cumulant, irreducible part of 
the fourth order correlator (for the sixth order
correlator it has been neglected). Note that,
for a gaussian process $D_{1,2,3,4}$ is exactly zero.
Therefore, in order to describe departures from 
gaussian behavior we consider an evolution equation 
for $D_{1,2,3,4}$. This can be accomplished starting with 
the evolution equation for the free waves (the
Zakharov equation) and developing evolution equations 
for the second and fourth moment of $a({\bf k})$. 
the equation for $D_{1,2,3,4}$
can be solved supposing that the wave action
density spectrum 
changes on a slow time scale and 
that at time $t=0$ the
waves are Gaussian (the cumulant is exactly zero).
Details can be found in \cite{JAN03}.
The result is the following equation for $D_{1,2,3,4}$:
\begin{equation} \label{CUM}
\begin{split}
D_{1,2,3,4}=&2 T_{1,2,3,4} \delta_{1+2-3-4}G(\Delta\omega,t) \\
&N_1N_2(N_3+N_4)-(N_1+N_2) N_3 N_4
\end{split}
\end{equation}
$T_{1,2,3,4}$ is the coupling coefficient for
the Zakharov equation and $G(\Delta \omega,t)$ is given by:
\begin{equation}
G(\Delta \omega,t)=
\frac{1-Cos(\Delta \omega t)} {\Delta \omega}
+i \frac{Sin(\Delta \omega t)} {\Delta \omega}
\end{equation}
with $\Delta \omega = \omega_1+\omega_2-\omega_3-\omega_4$.
We now consider the statistical properties of the weakly 
nonlinear system described by the Hamiltonian in 
(\ref{HAMIL}).
We consider the third order moment $<\eta({\bf x})^3>$; using the 
definition of the Fourier transform and equation 
(\ref{SURF}) we obtain:
\begin{equation} \label{skew0}
\begin{split}
&<\eta({\bf x})^3> = \int L_{1,2,3} (<a_1 a_2 a_3>+\\
&+<a_1^* a_2^* a_3^*>)\delta_{123} dk_{123} 
+3  \int L_{1,2,3}  (<a_1 a_2 a_3^*>+
\\ & +<a_1^* a_2^* a_3>)\delta_{1+2-3} dk_{123} 
\end{split}
\end{equation}
where 
\begin{equation}
L_{1,2,3}=\sqrt {\frac{\omega_1 \omega_2 \omega_3} {8 g^3}}
\end{equation}
Note that $L_{1,2,3}$ is symmetric under the transposition of 
all subscripts. We now use the canonical
transformation (\ref{CAN}) and insert it in (\ref{skew0})
to obtain:
\begin{equation} \label{skew1}
\begin{split}
<\eta({\bf x})^3>=&\int S_{1,2,3,4} 
(<b_1 b_2 b_3^* b_4^*>+ \\
&<b_1^* b_2^* b_3 b_4>) \delta_{1+2-3-4} dk_{1234}
\end{split}
\end{equation}
where the kernel $S_{1,2,3,4}$ is given in 
\cite{ONO05} where
also intermediate steps in the derivation are
reported. Here we just mention that 
$S_{1,2,3,4}$ is a function of $A^{(1)}$, $A^{(2)}$ and 
$A^{(3)}$ in the canonical trnasformation.
Now using equations (\ref{CORRE4}) and (\ref{CUM}) we
can write the skewness as:
\begin{equation} \label{skewf}
\begin{split}
&<\eta({\bf x})^3> =
4 \int  S_{1,2,1,2} N_1 N_2 dk_{12}+\\
&+ 16 \int S_{1,2,3,4} T_{1,2,3,4} N_1 N_2 N_3 
\frac{1-Cos(\Delta \omega t)} {\Delta \omega} \\
&\delta_{1+2-3-4} dk_{1234}
\end{split}
\end{equation}
The final result is that 
the skewness can be written as the sum of two contributions:
 the first one includes only 
bound modes and the second one depends both 
on bound and free modes. 

We now consider the kurtosis. Using a similar approach
used to derive equation (\ref{skewf}), we write the kurtosis in 
the following way:
\begin{equation} \label{kurt0}
\begin{split}
<\eta({\bf x})^4>= &
 \int M_{1,2,3,4} (<a_1 a_2 a^*_3 a^*_4>+\\
&+ <a_1^* a_2^* a_3 a_4>)\delta_{1+2-3-4} dk_{1234}
\end{split}
\end{equation}
where 
\begin{equation} \label{kurtcoef}
M_{1,2,3,4}=\frac{3} {4 g^2} \sqrt{\omega_1 \omega_2 \omega_3 \omega_4}
\end{equation}
Using the canonical transformation, we obtain
\begin{equation} \label{kurt1}
\begin{split}
&<\eta({\bf x})^4>=  \int M_{1,2,3,4} (<b_1 b_2 b^*_3 b^*_4>+ \\
& +<b_1 b_2 b^*_3 b^*_4> )\delta_{1+2-3-4} dk_{1234} +\\
& \int K_{1,2,3,4,5,6}  (<b_1 b_2 b_3 b_4^* b_5^* b_6^*>+ \\
& <b_1 b_2 b_3 b_4^* b_5^* b_6^*>) \delta_{1+2+3-4-5-6} dk_{123456} 
\end{split}
\end{equation}

The kernel $K_{1,2,3,4,5,6}$, given in \cite{ONO05},
is a function of  $A^{(1)}$, $A^{(2)}$, 
$A^{(3)}$. Note that at the same order, higher order terms
that we have omitted in the canonical transformation 
should enter (those terms are
reported in \cite{ONO05}).
We now apply the quasi- gaussian approximation and
use equations (\ref{CORRE4}) and (\ref{CORRE6}) 
for the fourth and sixth order correlator to obtain:
\begin{equation} \label{kurtf}
\begin{split}
 & <\eta({\bf x})^4>=
3 <\eta({\bf x})^2>^2 + \\
& + 16 \int M_{1,2,3,4} T_{1,2,3,4} N_1 N_2 N_3 
\frac{1-Cos(\Delta \omega t)} {\Delta \omega} \\
&\delta_{1+2-3-4} dk_{1234}  + 12  \int K_{1,2,3,1,2,3} N_1 N_2 N_3 dk_{123}
\end{split}
\end{equation}
This is the final formula for the kurtosis expressed
as a sum of 3 contributions: the first one corresponds
to a pure gaussian system, the second one
is the contribution from free modes
and the last one is the contribution from 
bound modes. 
Equations (\ref{skewf}) and (\ref{kurtf})
are valid for long and short crested waves.
It is interesting to note that in the limit of 
long-crested waves and in the narrow-banded
approximation the integral including free modes
becomes proportional to the Benjamin-Feir 
Index (see  \cite{JAN03} for
details), while the integral including bound modes
become proportional to the steepness squared.

As a conclusion we may state that
the main contribution of 
the paper is the derivation of equations 
(\ref{skewf}) and (\ref{kurtf}). Those equations,
 obtained using the Hamiltonian 
formulation of surface gravity waves,
can be viewed as a generalization and unification 
of the theory developed by Longuet-Higgins in 1963 on bound modes
and  the theory developed by Janssen in 2003
on free modes.

{\bf Acknowledgments}
The experimental  work at Marintek (Norway)
 has been supported  by 
the Improving Human Potential - Transnational 
Access to Research Infrastructures Programme 
of the European Commission
under the contract HPRI-CT-2001-00176. 
This research has also been supported 
by the U.S. Army Engineer Research and Development Center.
MIUR is also acknowledged.
%


\end{document}